\documentclass[fleqn,usenatbib]{mnras}

\usepackage{newtxtext,newtxmath}
\usepackage{xcolor}
\usepackage{hyperref}
\usepackage{academicons}

\usepackage[T1]{fontenc}

\DeclareRobustCommand{\VAN}[3]{#2}
\let\VANthebibliography\thebibliography
\def\thebibliography{\DeclareRobustCommand{\VAN}[3]{##3}\VANthebibliography}

\usepackage{graphicx}	
\usepackage{amsmath}	

\title[Statistical Bias in $H_0$ for Mock Lenses]{Statistical Bias in the Hubble Constant and Mass Power Law Slope for Mock Strong Lenses}

\author[Ruan \& Keeton]{
Dilys Ruan,$^{1}$\thanks{E-mail: druan@physics.rutgers.edu}
Charles R. Keeton$^{1}$ \\
$^{1}$ Department of Physics and Astronomy, Rutgers University, Piscataway, NJ 08854, USA
}

\date{Accepted XXX. Received YYY; in original form ZZZ}

\pubyear{2023}

\begin{document}
\label{firstpage}
\pagerange{\pageref{firstpage}--\pageref{lastpage}}
\maketitle

\begin{abstract}
Strong gravitational lensing offers constraints on the Hubble constant that are independent of other methods. However, those constraints are subject to uncertainties in lens models. Previous studies suggest that using an elliptical power law + external shear (EPL+XS) for the lensing galaxy can yield results that are precise but inaccurate. We examine such models by generating and fitting mock lenses which produces multiple images of a background quasar-like point source. Despite using the same model for input and output, we find statistical bias in the Hubble constant on the order of 3\% to 5\%, depending on whether the elliptical lenses have noise or not. The phase space distribution has a `flared' shape that causes the mass power law slope to be underestimated and the Hubble constant to be overestimated. The bias varies with image configuration, which we quantify through annulus length between images with the first and second time delays ($\Delta r_{1,2}$). The statistical bias is worse for configurations that have narrow annuli (e.g., symmetric cross configurations). Assuming a source at redshift 2.0 and an EPL+XS lens at redshift 0.3, we find that the bias can be reduced, but not eliminated, if we limit the sample to systems with annulus lengths $\Delta r_{1,2} \gtrsim 0.3$ arcsec. As lens samples grow, it may be helpful to prioritize this range of image configurations for follow-up observation and analysis. 
\end{abstract}

\begin{keywords}
gravitational lensing: strong -- (cosmology:) cosmological parameters -- galaxies: general
\end{keywords}



\section{Introduction}

The Hubble tension describes a 4-6$\sigma$ discrepancy between `local' and `distant' measurements of the current expansion rate of the Universe, or the Hubble constant ($H_0$). Local measurements such as those from Cepheid stars + Type Ia supernovae \citep{Riess2019,Riess-et-al-2021}, Tip of the Red Giant Branch stars + Type Ia supernovae \citep{Freedman2019,Freedman-2021}, gravitational lensing \citep{H0LiCOWXIII,TDCOSMOIV}, and masers \citep{Reid2019,Pesce-2020} tend to find a higher value of $H_0$ than the distant measurements made with the cosmic microwave background (CMB) and baryon acoustic oscillations \citep{Planck2018,Planck2020}. The tension could be interpreted as evidence for a cosmological model that deviates from fiducial $\Lambda$CDM \citep{Knox-Millea-2020,Hubble2021}, but before drawing such a conclusion it is crucial to assess possible sources of errors with each measurement. In this study we examine constraints on $H_0$ from gravitational lensing.

Gravitational lensing provides a local measurement of $H_0$ that is independent of the distance ladder. In strong lensing, the gravitational deflection of light creates multiple images of a background source, and any variability in the source will be delayed between images because the light rays travel different paths. Often the background source is a quasar or supernova, while the foreground lens is a massive elliptical galaxy. The COSMOGRAIL team has undertaken long-term monitoring to measure lensing time delays (even accounting for additional variability produced by microlensing; \citealt{COSMOGRAIL2018, COSMOGRAIL}), and upcoming surveys such as the Rubin Observatory's Legacy Survey of Space and Time (LSST) should significantly increase the sample \citep{2012LSST, 2010Oguri}.

Time delays must be interpreted in the context of a lens model, and this is where statistical and systematic errors may arise. Most lens models are categorized by their treatment of radial and angular components in the lens potential. Objects in the environment of the main lens galaxy or along the line-of-sight also contribute to the lens potential, and these effects are often treated with a multipole expansion. A common simplification in lens modelling is to fit the main lens galaxy with an elliptical power law (EPL) and then add external shear (XS) to account for environmental and line-of-sight effects. However, there are some exact or approximate degeneracies in lens models, such as the mass sheet degeneracy and radial profile degeneracy \citep{Falco1985, Saha2000, Kochanek2002}. Stellar velocity dispersions provide a lensing-independent way to break these degeneracies and constrain the density profile \citep[e.g.,][]{1999Romanowsky, 2002Treu, H0LiCOWXIII, TDCOSMOIV}. An isothermal elliptical model appears to be fairly accurate near the Einstein radius where images form, but this is not necessarily true at smaller or larger radii \citep{2009vandeVan, 2013Cappellari}.

After analyzing six lens systems from COSMOGRAIL, the H0LiCOW team measured $H_0 = 73.3_{-1.8}^{+1.7}$ km s$^{-1}$ Mpc$^{-1}$ \citep{H0LiCOWXIII}. The TDCOSMO team used hierarchical Bayesian analysis on these six lenses plus one more system to obtain $H_0=74.5^{+5.6}_{-6.1}$ km s$^{-1}$ Mpc$^{-1}$ \citep{TDCOSMOIV}. To improve the measurement, they included another 33 systems from SLACS, nine of which have resolved stellar kinematics. Assuming that galaxies from SLACS and TDCOSMO are drawn from the same parent population, the team used joint Bayesian analysis to find $H_0 = 67.4_{-3.2}^{+4.1}$ km s$^{-1}$ Mpc$^{-1}$ -- a 5\% precision measurement that is closer to the $H_0$ value measured from the CMB. 

The sample of strong lenses will grow by orders of magnitude with current and upcoming surveys. STRIDES is expected to find about 250 lens candidates in Stage 2 \citep{2018STRIDES}, and has recently discovered 30 new lenses \citep{2023STRIDES}. The Rubin Observatory's LSST should discover more than 8000 lenses through high-cadence, 10-year observations of 18000 deg$^2$ of the sky \citep{2012LSST, 2010Oguri}. Euclid, Rubin, the Square Kilometre Array, and JWST will collectively provide deep, multi-wavelength observations of $\sim\!10^5$ lenses \citep{2015SKA, 2020Weiner}. As larger samples improve the statistical precision, it is ever more important to understand systematic uncertainties for competitive $H_0$ measurements \citep{2021TDLMC}. 

Multiple studies have examined bias in $H_0$ caused by mismatches between true and fitted mass distributions. Many studies have generated lenses with a composite model (i.e., baryons + dark matter) with XS, and fitted the time delay data with an EPL+XS. \citet{Schneider2013} fitted each lens with its true stellar velocity dispersion. Although best-fit models reproduced the data well, $H_0$ bias was $\sim\!20\%$. \citet{Gomer-Williams-slope} similarly found that kinematic constraints from a Jeans analysis led to $H_0$ bias ranging from 0-23\%, depending on the image radius. \citet{Kochanek2020} noted that if few-parameter mass distributions (like the EPL+XS) are assumed, it is possible to obtain results that are precise but inaccurate. In contrast to this result, \citet{2020A&A...639A.101M} explored systematic uncertainties in the TDCOSMO analysis and found that composite and EPL models with XS gave statistically consistent results.

In this study, we identify a statistical bias that arises even if the \emph{same} mass distribution is used to generate and fit mock lens galaxies. We consider three classes of models: a circular power law that offers some insightful analytic results, as well as circular and elliptical power laws with external shear. Section \ref{methods} describes our methodology, Section \ref{results} presents our results for the three model classes, and Section \ref{conclusions} summarizes our conclusions. Appendix \ref{app-choices} examines some technical aspects of our modelling analysis. We assume a flat $\Lambda$CDM cosmology, with $\Omega_m=0.315$, $\Omega_\Lambda=0.685$, and $h=0.7$ where $H_0=100h$ km s$^{-1}$ Mpc$^{-1}$. Code is available on GitHub to demonstrate how we performed the analysis \footnote{\url{https://github.com/dilysruan/paper-ruan-keeton-2023}}.

\section{Methods}
\label{methods}
\subsection{Time Delay Cosmography \& Lens Model}
\label{models}

We summarize key aspects of time delay cosmography and refer the reader to comprehensive reviews for more details on this topic \citep[e.g.,][]{Treu2016, 2022Shajib} and gravitational lensing more broadly \citep[e.g.,][]{Congdon2018}. According to Fermat's Principle, light rays follow paths that correspond to stationary points of the time delay surface
\begin{equation}
  \tau(\boldsymbol{\theta}|\boldsymbol{\beta}) = \frac{D_{\Delta t}}{c} \Phi(\boldsymbol{\theta}|\boldsymbol{\beta})\text{, }
\label{timedelay}
\end{equation}
where $\boldsymbol{\theta}$ is the angular position in the image plane, $\boldsymbol{\beta}$ is the intrinsic angular position of the source, and the time delay distance $D_{\Delta t}$ depends on the lens and source redshifts through cosmological angular diameter distances,
\begin{equation}
    D_{\Delta t} = (1+z_{l}) \frac{D(z_l)\,D(z_s)}{D(z_l,z_s)}\text{.}
    \label{timedelaydist}
\end{equation}
The Fermat Potential is
\begin{equation}
    \Phi(\boldsymbol{\theta}, \boldsymbol{\beta}) = \frac{1}{2} \left( \boldsymbol{\theta} - \boldsymbol{\beta}\right)^2 - \psi(\boldsymbol{\theta})\text{,}
    \label{len_potential}
\end{equation}
where $\psi(\boldsymbol{\theta})$ is the projected gravitational potential of the lens. The first term in $\Phi$ accounts for the travel time due to the geometric path, while the second term in $\Phi$ accounts for relativistic time dilation. The basic idea of time delay cosmography is to measure the time delay $\Delta\tau_{AB}$ between two images, use a lens model to compute the difference in Fermat potential $\Delta\Phi_{AB}$, and then use Equation (\ref{timedelay}) to solve for the time delay distance and constrain $H_0$.

The source and image positions are related by the lens equation 
\begin{equation}
    \boldsymbol{\beta} = \boldsymbol{\theta} - \boldsymbol{\alpha(\boldsymbol{\theta})}\text{,}
\end{equation}
where the deflection angle $\boldsymbol{\alpha} = \nabla \psi(\boldsymbol{\theta})$ is the gradient of the lens potential. The lens potential is determined by the surface mass density through the Poisson equation
\begin{equation}
    \nabla^2 \psi(\boldsymbol{\theta}) = 2 \kappa(\boldsymbol{\theta})
    = \frac{2 \Sigma(\boldsymbol{\theta})}{\Sigma_{crit}} \text{,}
\end{equation}
where the convergence $\kappa$ is the surface mass density scaled by the critical density for lensing,
\begin{equation}
  \Sigma_{crit} = \frac{c^2}{4\pi G}\,\frac{D(z_s)}{D(z_l) D(z_l,z_s)} \text{.}
\end{equation}
The lensing magnification is given by $\mu = \det\boldsymbol{A}$ where the amplification tensor is $\boldsymbol{A} = {\partial \boldsymbol{\theta}}/{\partial \boldsymbol{\beta}} = ({\partial \boldsymbol{\beta}}/{\partial \boldsymbol{\theta}})^{-1}$.

We consider a power law density profile. For circular symmetry, the lens potential, deflection angle, and convergence have the respective forms
\begin{align}
    \psi(r) &= \frac{R_E^{2-\eta}}{\eta} r^\eta \text{, } \label{EPL}\\
    \alpha(r) &= R_E^{2-\eta} r^{\eta-1}\text{, } \\
    \kappa(r) &= \frac{\eta}{2} R_E^{2-\eta} r^{\eta-2} \text{.}
\end{align}
Here $R_E$ is the Einstein radius and $\eta$ is the mass power law slope. An isothermal profile has $\eta=1$. It can be convenient to express the normalization through a simple multiplicative factor\footnote{The goodness of fit $\chi^2_{pos}$ in Equation (\ref{chi_pos}) is quadratic in $b$.}
\begin{equation}
    b = R_E^{2-\eta}\text{,}
    \label{bmode}
\end{equation}
and we will use this form for the majority of our analysis.
We can obtain an elliptical mass distribution by replacing $r$ in the convergence with the elliptical radius $\sqrt{q^2 x^2 + y^2}$, where $0 < q \leq 1$ is the ratio of the short to long axes. In models we use the quasi-Cartesian components that are expressed in terms of the ellipticity $e=1-q$ and orientation angle $\theta_e$ as $e_c = e \cos 2\theta_e$ and $e_s = e \sin 2\theta_e$. We implement expressions for the potential, deflection, and magnification of the elliptical power law model from \citet{2015Tessore-Metcalf,2016Tessore-Metcalf}. In some cases we also consider an external shear, whose lens potential has the form
\begin{equation}
  \psi_{XS}(\boldsymbol{\theta}) = \frac{1}{2} \boldsymbol{\theta} \cdot \boldsymbol{\Gamma} \cdot \boldsymbol{\theta}
  \quad\mbox{where}\quad
  \boldsymbol{\Gamma} = \left[\begin{array}{cc}
    \gamma_c & \gamma_s \\
    \gamma_s & -\gamma_c 
  \end{array}\right] \text{.}
\end{equation}
Here $\gamma_c$ and $\gamma_s$ are quasi-Cartesian components of shear, which are related to the shear amplitude $\gamma$ and direction $\theta_\gamma$ as $\gamma_c = \gamma \cos 2\theta_\gamma$ and $\gamma_s = \gamma \sin 2\theta_\gamma$.

\subsection{Generating Mock Data}
\label{mock-data}

\begin{figure}
    \centering
    \includegraphics[width=0.3\textwidth]{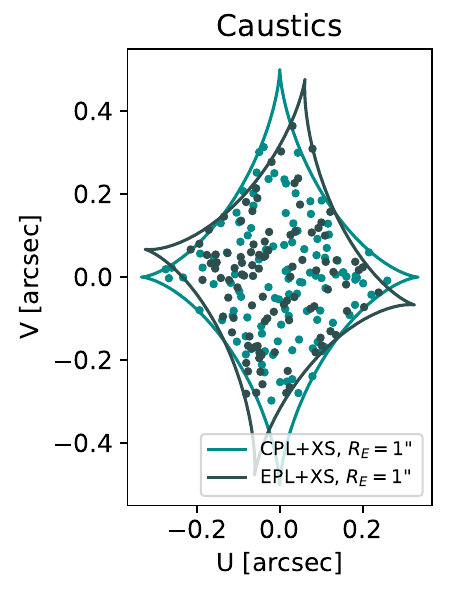}
    \caption{Caustic curves for the CPL+XS {(teal)} and EPL+XS {(dark teal)} lenses. There are 100 sources {(markers)} at different positions within the caustic curves for each mass distribution.}
    \label{caus}
\end{figure}

We use \verb|pygravlens|\footnote{\url{https://github.com/chuckkeeton/pygravlens}}, a python version of the gravitational lensing code \verb|gravlens| \citep{lensmodel}. We choose as our fiducial model an isothermal mass distribution ($\eta=1$) with a lens galaxy at $z_l = 0.3$ and a source quasar 
at $z_s = 2.0$. We create circular models with and without shear, and an elliptical model with shear. The fiducial parameter values are given in Table \ref{truth}. For each mass distribution, we choose random source positions from a uniform distribution that is centered on the origin and spans the caustic curve. Both the CPL+XS and EPL+XS caustics span $\sim 0.8$ arcsec along the long axis, as shown in Figure \ref{caus}. 

\begin{table*}
    {
    \begin{tabular}{ccccccccc}
        \hline
        Case & $R_E$ & $e_c$ & $e_s$ & $\gamma_c$ & $\gamma_s$ & $\eta$ & $h$ & $\sigma_\tau$ [days] \\
        \hline
        Doubly-Imaged CPL & 1.0 & 0 & 0 & 0 & 0  & 1.0 & 0.7 & 3.0 \\
        Quadruply-Imaged CPL+XS & 1.0 & 0 & 0 & 0.2 & 0  & 1.0 & 0.7 & 2.0, 4.0 \\
        Quadruply-Imaged EPL+XS & 1.0 & 0.15 & 0.15 & 0.22 & 0  & 1.0 & 0.7 & 2.0 \\
        \hline
    \end{tabular}
    \par }
    \caption{
    Input values for the various lens model parameters.
    $R_E$ is the Einstein radius, $e_c$ and $e_s$ are quasi-Cartesian components of ellipticity, and $\gamma_c$ and $\gamma_s$ are the corresponding quasi-Cartesian components of external shear. The input $\Lambda$CDM cosmology has $\Omega_m=0.315$, $\Omega_\Lambda=0.685$, and $h=0.7$.
    }
    \label{truth}
\end{table*}

\begin{figure}
    \centering
    \includegraphics[width=0.4\textwidth]{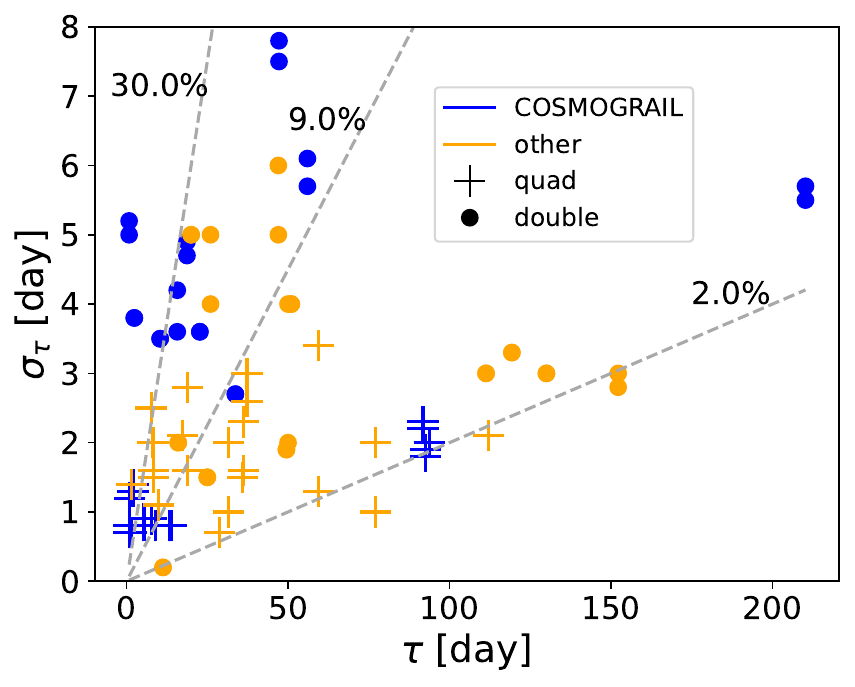}
    \caption{
    The relative time delay values ($\tau$) and uncertainties ($\sigma_\tau$) for observed lens systems listed in Table 3 of \citet{COSMOGRAIL}. The dashed lines indicate certain percentage errors.
    }
    \label{delay-compare}
\end{figure}

The image position uncertainties are 0.003 arcsec, and flux uncertainties are 5\%. To identify reasonable values for the time delay uncertainties, Figure \ref{delay-compare} shows observed time delays and uncertainties from Table 3 in \citet{COSMOGRAIL}, which includes values from the COSMOGRAIL study and previous studies \citep{2018Biggs_Brown, 2018Courbin, 2013Hainline, 2005Jakobsson, 2013Rathna, 2007Poindexter, 2018Bonvin, 2013Eulaers, 2016Goicoechea, 2017Akhunov, 2001Patnaik, 2019Shalyapin, 2002Burud, 2000Burud, 2002Fassnacht, 2007Vuissoz, 1998Lovell, 2019Bonvin}.
Time delays can range from a few days to 100 days or more, and there is no obvious trend in terms of percentage uncertainty. We therefore choose to use the median values of the uncertainties, which are $3.0$ days for doubly-imaged systems and $2.0$ days for quadruply-imaged systems. We adopt these as the time delay uncertainties for our mock lenses. Statistical noise is only added to the mock data in Section \ref{noise}.

\subsection{Fitting Models}
\label{algorithm}

When fitting models, we can impose constraints from image positions, flux ratios, and time delays through the following goodness-of-fit functions:
\begin{align}
    \chi_{pos}^2 &= \sum_{i} \left[ \boldsymbol{\beta} + \boldsymbol{\alpha}_i - \boldsymbol{\theta}_i\right]^T \boldsymbol{A}_i^T \boldsymbol{C}_i^{-1} \boldsymbol{A}_i \left[ \boldsymbol{\beta} + \boldsymbol{\alpha}_i - \boldsymbol{\theta}_i\right]\text{,} \label{chi_pos} \\
    \chi_{flux}^2 &= \sum_{i} \frac{\left( \mu_i f_{src} - f_{obs,i}\right)^2}{\sigma^2_{f,i}}
    \text{, }  \label{chi_flux} \\
    \chi_{tdel}^2 &= \sum_{i} \frac{\left( \Delta t_{mod,i} - \Delta t_{obs,i}\right)^2}{\sigma_{t,i}^2}\text{.} \label{chi_tdel}
\end{align}
Here the observed quantities are the image positions $\boldsymbol{\theta}_i$ and covariance matrices $\boldsymbol{C}_i$ for the position uncertainties, the fluxes $f_{obs,i}$ and their uncertainties $\sigma_{f,i}$, and the time delays $\Delta t_{obs,i}$ and their uncertainties $\sigma_{t,i}$.
The model quantities are the source position $\boldsymbol{\beta}$ and source flux $f_{src}$, as well as parameters for the mass model ($R_E$ or $b$ and $\eta$, along with ellipticity and/or shear if they are used). The Hubble parameter $h$ enters the predicted time delays through the scaling $\Delta t_{mod} \propto h^{-1}$. 

In some cases we also impose a prior on the power law slope through
\begin{equation}
    \chi_{PL}^2 =  \frac{\left( \eta_{mod} - \eta_{obs}\right)^2}{\sigma_{PL}^2}\text{.}  \label{chi_PL}
\end{equation}
This prior is meant to mimic adding constraints from stellar kinematics. Following constraints on the power law slope given in Table 2 of \citet{TDCOSMOIV}, we adopt $\eta_{obs} = 1.0$ and $\sigma_{PL} = 0.1$.

We constrain the model parameters using Bayesian inference with the likelihood function $L \propto e^{-\chi^2/2}$. We work analytically in certain cases described below. More generally, we sample the Bayesian posterior using MCMC methods implemented in the code \verb|emcee| \citep{emcee}. We assume flat priors on the parameters (except for $\eta$ as noted above).
In some cases the parameters are well constrained such that we do not need to impose explicit bounds. However, other cases are not well constrained, so we impose the bounds $\eta \in (0.5,1.5)$ and $h \in (0,1.2)$ in order to avoid values that are implausible or unphysical. MCMC runs for the CPL+XS and EPL+XS systems use 100 walkers, 8000 burn-in steps, and 8000 main steps.

There are several choices that must be made in the modelling process. Nuisance parameters can be fully marginalized or merely optimized (i.e., set to their best-fit values). For parameters that are quadratic in $\chi^2$, the optimal values can be computed analytically to reduce computational effort. Statistically, however, marginalization is preferred. As noted above, the mass model can be specified using either the Einstein radius $R_E$ or the mass parameter $b$. Our fiducial results use marginalization and the `b mode' normalization. Flux ratio constraints are used for the CPL case in order to break the degeneracy for doubly-imaged systems. We do not apply them for the quadruply-imaged CPL+XS or EPL+XS cases since flux ratios can be affected by microlensing and are not generally used as constraints in time delay studies \citep{2019Chen, 2020Rusu}.  Appendix \ref{app-choices} compares other modelling choices and shows that they do not significantly change our results.

\section{Results \& Discussion}
\label{results}

\subsection{Doubly-Imaged Circular Power Law}
\label{doub-circ}

We begin with the simple case of a circular lens (with no external shear) so key results can be understood analytically. A circular lens can produce two images on opposite sides of the origin; let their distances from the origin be $\theta_A > \theta_B > 0$. Then the lens equations for the two images can be written as
\begin{align}
    \beta &= \theta_A - b\,\theta_A^{\eta-1} \\
    -\beta &= \theta_B - b\,\theta_B^{\eta-1} \nonumber
\end{align}
The equation for image B has a minus sign on the left because the image lies on the opposite side of the lens from the source. These two equations can be solved for $b$ and $\beta$,
\begin{equation}\label{CPL-b-beta}
    b = \frac{\theta_A + \theta_B}{\theta_A^{\eta-1} + \theta_B^{\eta-1}}
    \quad\mbox{and}\quad
    \beta = \frac{\theta_A \theta_B^{\eta-1} - \theta_B \theta_A^{\eta-1}}{\theta_A^{\eta-1} + \theta_B^{\eta-1}}
\end{equation}
which means the image positions can be reproduced by \emph{any} value of the power law slope $\eta$. Thus, flux constraints are needed to break the radial profile degeneracy. Using the analytic solutions for $b$ and $\beta$ ensures that the goodness of fit for the image positions is $\chi^2_{pos} = 0$, so the overall goodness of fit in this analysis is $\chi^2 = \chi_{flux}^2 + \chi_{tdel}^2$.

\begin{figure}
    \centering
    \includegraphics[width=0.35\textwidth]{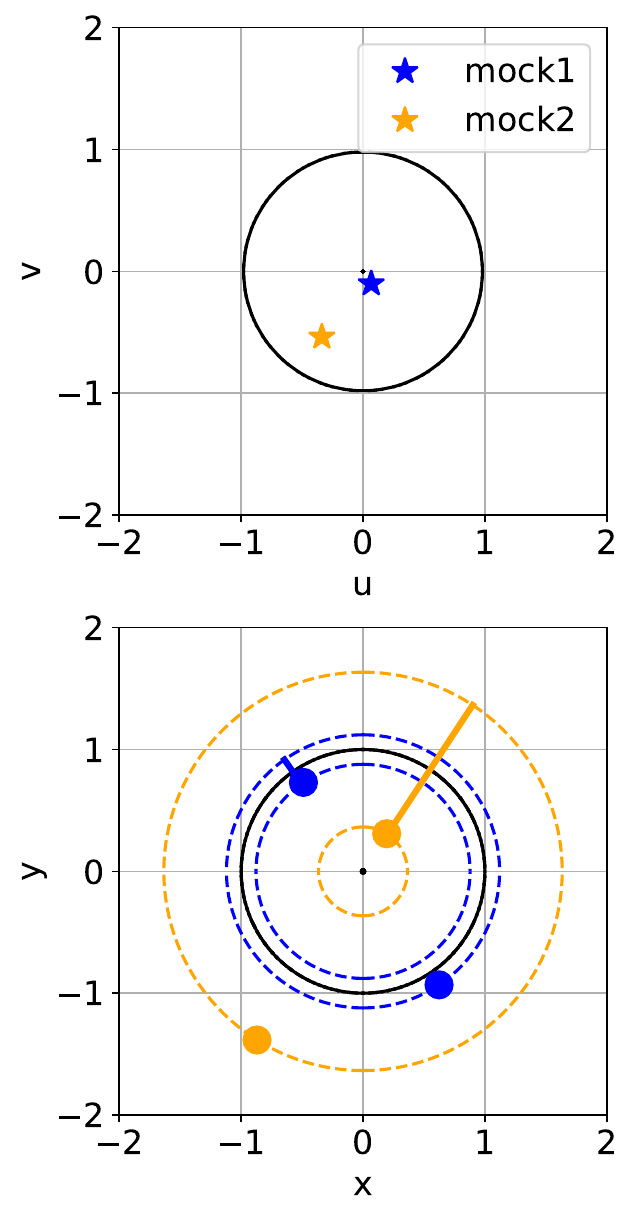}
    \caption{
    The source plane \emph{(top)} and image plane \emph{(bottom)} for the doubly-imaged circular lens systems CPL Mock 1 (blue markers) and CPL Mock 2 (orange markers). The caustic and critical curves are shown with the solid black lines. Mock 1 has a source closer to the origin, which results in images that are closer together in radii, while the images from Mock 2 are more spread out. The annulus made by each lens system are shown with the dashed lines. The solid colored line represents the annulus size between images for each lens system. Axes are labeled in units of arcsec.
    }
    \label{doub_circ1&2}
\end{figure}

We consider two mock systems with different source locations, as shown in Figure \ref{doub_circ1&2}. The source for Mock 1 is closer to the origin, so the two images are closer to the Einstein radius and have higher magnifications. The annulus spanned by the images is narrower for Mock 1 and wider for Mock 2.

\begin{figure}
    \centering
    \includegraphics[width=0.43\textwidth]{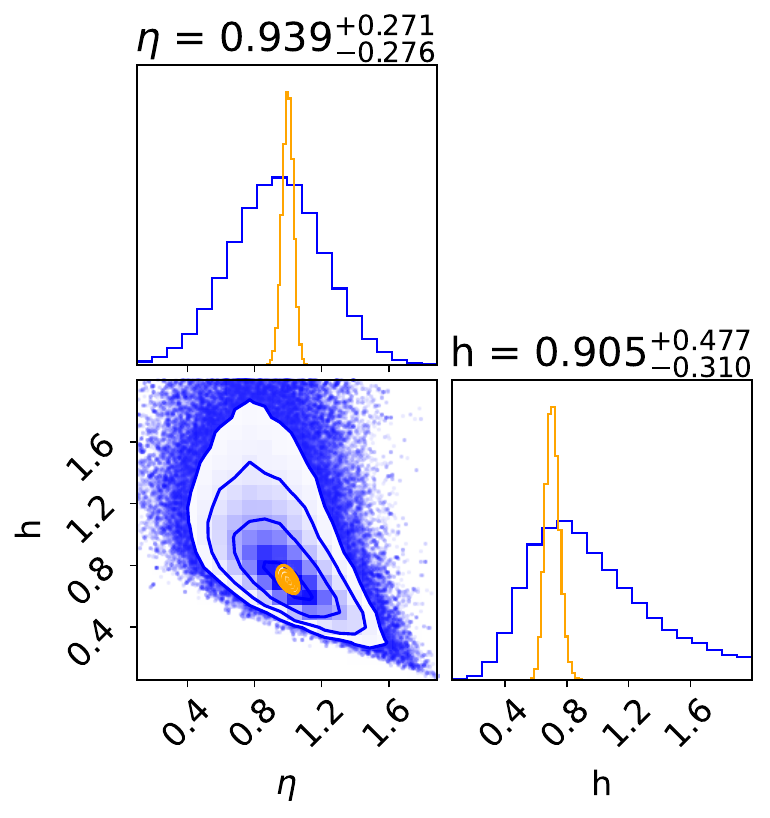}
    \caption{
    MCMC results for CPL Mock 1 {(blue)} and Mock 2 {(orange)}. Column titles give the median values and 68\% confidence intervals for Mock 1. For Mock 2, the corresponding results are $\eta=0.999^{+0.033}_{-0.034}$ and $h=0.707^{+0.048}_{-0.043}$.
    }
    \label{doub-circ-corner}
\end{figure}

Since we solve for $b$ and $\beta$ using Equation (\ref{CPL-b-beta}), and we optimize the intrinsic flux of the source (which can also be done analytically since $\chi^2_{flux}$ is quadratic in $f_{src}$), the only parameters that need to be explored by MCMC are $\eta$ and $h$. Figure \ref{doub-circ-corner} shows posterior distributions for both Mock 1 and Mock 2. The posterior for Mock 2 is quite narrow and encloses the input values, with medians of $\eta=0.999^{+0.033}_{-0.034}$ and $h=0.707^{+0.048}_{-0.043}$. By contrast, the posterior for Mock 1 is much broader. Although it encloses the input values, the medians are significantly biased with $\eta=0.939^{+0.271}_{-0.276}$ and $h=0.905^{+0.477}_{-0.310}$.

\begin{figure}
    \centering
    \includegraphics[width=0.4\textwidth]{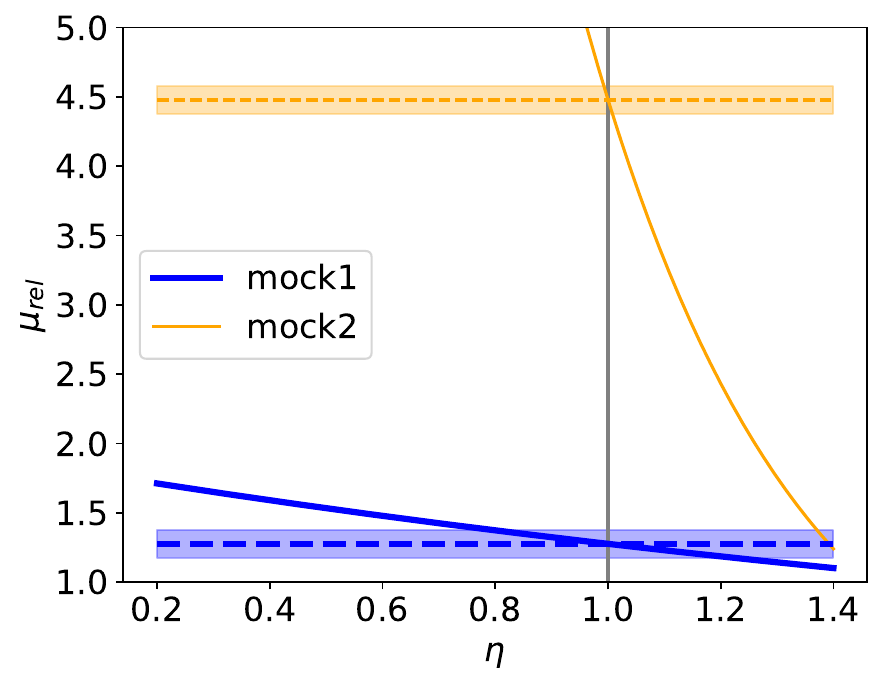}
    \caption{
    The solid curves show the predicted image magnification ratio as a function of the power law slope for Mock 1 {(thicker blue)} and Mock 2 {(orange)}. The dashed lines and shaded regions represent the `observed' magnification ratios and uncertainties for the two systems. The grey vertical line indicates the input power law slope, $\eta=1$.
    }
    \label{doub-circ-mu}
\end{figure}

The difference in posteriors can be understood using Figure \ref{doub-circ-mu}, which shows the magnification ratio of the images as a function of the power law slope $\eta$. For Mock 2, the curve is quite steep because the image radii are different; as a result, only a narrow range of $\eta$ values can produce a magnification ratio that is consistent with the `observed' value. For Mock 1, by contrast, the curve of magnification ratio vs.\ $\eta$ is shallower because the image radii are similar; in this case, the range of $\eta$ values that are consistent with the data is much broader. In other words, the narrow image annulus in Mock 1 provides worse sampling of the mass profile, leading to larger uncertainties in the lens model.

\begin{figure}
    \centering
    \includegraphics[width=0.35\textwidth]{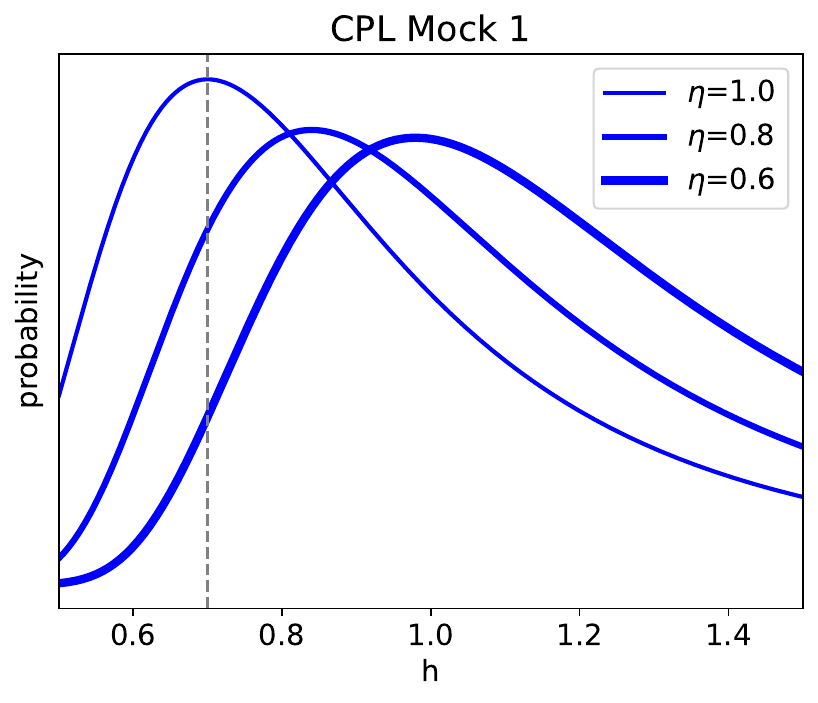}
    \caption{
    Analytic results for the conditional probability distribution $p(h|\eta)$ for CPL Mock 1 as the power law slope $\eta$ varies. As $\eta$ decreases, the curve not only shifts but also broadens. The grey dashed line shows the input truth $h=0.7$.
    }
    \label{hlike}
\end{figure}

It is important to note that the \emph{peak} of each posterior distribution corresponds to the input values of $\eta=1$ and $h=0.7$, because those values yield a perfect fit with $\chi^2=0$. However, the \emph{median} values of the marginalized distributions can be shifted away from the input values because of the shape of the posterior distribution. Figure \ref{doub-circ-corner} shows that the joint posterior $p(\eta,h)$ exhibits a `flared' shape: it is narrower in the lower right and becomes broader in the upper left. This effect can be understood analytically as follows. For each value of $\eta$, we can solve for $b$ and $\beta$ using Equation (\ref{CPL-b-beta}) and thus predict the model time delay $\Delta t_{mod}$. We can then compute the conditional probability distribution $p(h|\eta) \propto e^{-\chi^2_{tdel}}$ using Equation (\ref{chi_tdel}). Figure \ref{hlike} shows the resulting analytic curves for Mock 1. The scaling $\Delta t_{mod} \propto h^{-1}$ means that $h$ follows a reciprocal Gaussian distribution. As $\eta$ decreases, the $p(h|\eta)$ curve not only shifts but also broadens. The net effect is that there is more phase space at lower values of $\eta$ and higher values of $h$. Because of this statistical effect, the power law slope $\eta$ tends to be underestimated and $h$ tends to be overestimated. This statistical bias is larger in systems like Mock 1 that are less constrained and thus have larger model uncertainties.

\subsection{Quadruply-Imaged Circular Power Law + External Shear}
\label{quad-shear}

We expand the analysis to a more realistic model by including external shear. In this case the model can produce quad lenses, and we focus on them because having four images provides more constraints on the lens model. We pick random sources to produce 100 mock quad lenses. For each system, there are seven free parameters: two source position coordinates, Einstein radius, two components of shear, power law slope, and $h$.

Figure \ref{mock77-corner} shows the full MCMC results for CPL+XS Mock 77, a fold configuration. Here we use constraints from image positions and time delays, so the goodness of fit is $\chi^2 = \chi^2_{pos} + \chi^2_{tdel}$. 
The constraints from Mock 77 are strong enough to produce a single-peaked posterior distribution that is `compact' in the sense of being contained within the prior range, and the fitted value for each parameter is consistent with the input truth. Even so, the $(\eta,h)$ posterior shows some degree of `flaring' like we saw in the CPL case, leading to some bias in the median values.

Below we will see that EPL+XS lens models have more freedom and often need priors on the power law slope in order to be well constrained. To enable a fair comparison between CPL+XS and EPL+XS cases, we rerun the current circular models with the priors on $\eta$ such that the goodness of fit is now $\chi^2 = \chi^2_{pos} + \chi^2_{tdel} + \chi^2_{PL}$. To demonstrate the effect of having constraints on $\eta$, Figure \ref{CPLXS-corner} compares MCMC results for CPL+XS Mock 35, a highly symmetric cross configuration. The orange contours show the original results while the teal contours show results with the $\eta$ prior. The prior significantly reduces the phase space volume, but biases in $\eta$ and $h$ are still present.

\begin{figure}
    \centering
    \includegraphics[width=0.45\textwidth]{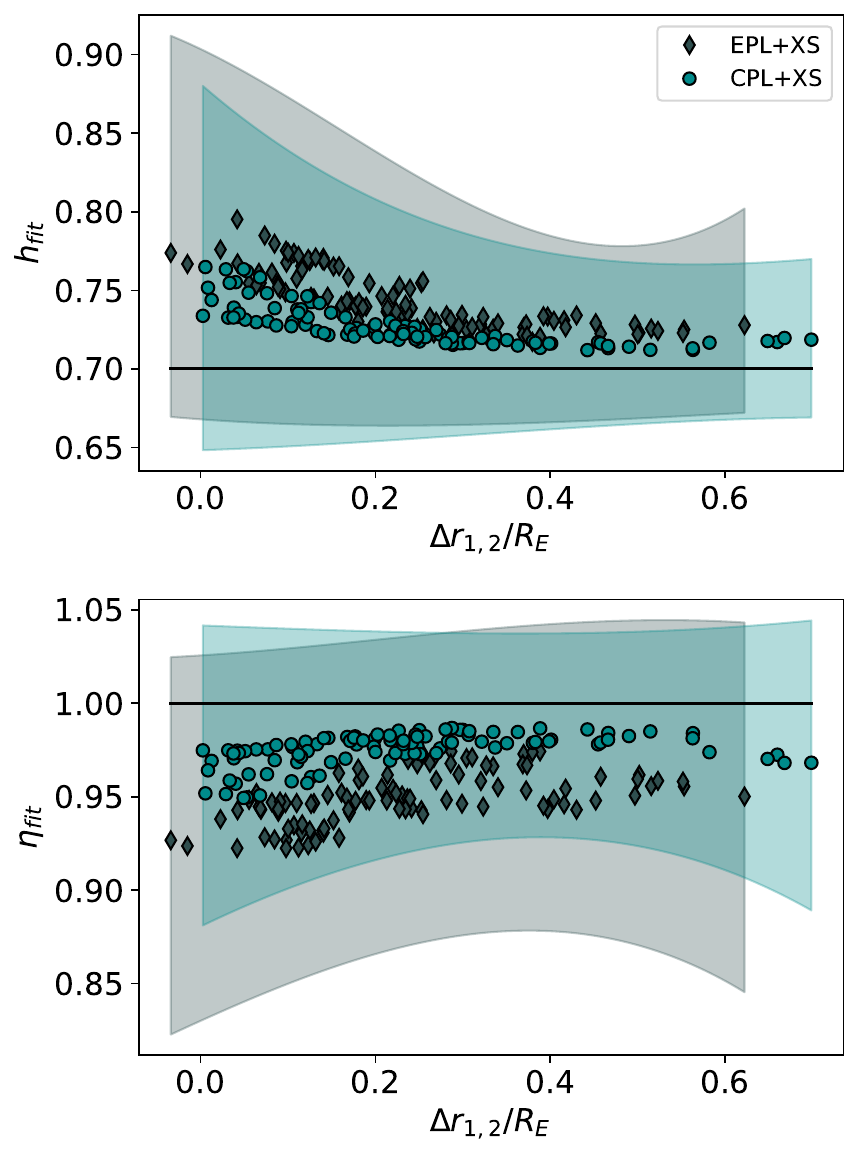}
    \caption{
    Results for $h$ \emph{(top)} and $\eta$ \emph{(bottom)} values compared to the annulus length normalized by $R_E$; points show median values while errorbars show 68\% confidence intervals. The CPL+XS case (teal) has less scatter than the EPL+XS case (dark teal). The input truth, $h=0.7$ and $\eta=1$, is shown with the black solid line. The shaded errors are interpolated with a univariate spline.
    }
    \label{hfit-annulus}
\end{figure}

We would like to understand how the biases in $\eta$ and $h$ vary with image configuration. Motivated by the circular double case, we consider the annulus spanned by the images, and we also consider time delays between different pairs of images. We find that the best predictors of biases are the time delay ($\tau_{1,2}$) or annulus length ($\Delta r_{1,2}$) between the first and second arriving images. We give preference to the annulus length because it is easily determined from initial observations of a lens. Figure \ref{hfit-annulus} shows the median values and uncertainties for $h$ and $\eta$ as a function of $\Delta r_{1,2}$ for all of the mock lenses in the CPL+XS case (teal).  Overall, $h$ is overestimated and $\eta$ is underestimated for each system. The bias and uncertainty are larger when the annulus is narrow (e.g., for symmetric configurations like crosses), and they decrease when the annulus is wider (e.g., for short-axis cusps and some folds).

\subsection{Quadruply-Imaged Elliptical Power Law + External Shear}
\label{quad-shear-ellip}

Since galaxies can be elliptical, we finally consider the EPL+XS model commonly used in the field. Ellipticity introduces two additional parameters, and this brings the total number of free parameters to nine. To provide enough constraints, we use realistic $\eta$ and $h$ bounds along with image positions, time delays, and power law slope constraints so that $\chi^2 = \chi^2_{pos} + \chi^2_{tdel} + \chi^2_{PL}$. 

Figure \ref{mock_ellip_maxbias} shows MCMC results for EPL+XS Mock 51, a cross configuration. The posterior is still single-peaked but slightly more complicated because there are more free parameters, and there are some covariances (such as between ellipticity and shear, which both describe the angular structure of the lens potential). Bias still exists for this cross configuration, likely due to its lack of constraints on the radial mass profile. Despite using the power law slope as a constraint and having physical ranges for $\eta$ and $h$, we find an inaccurate Hubble constant with $h=0.795^{+0.150}_{-0.127}$.

\begin{figure}
    \centering
    \includegraphics[width=0.35\textwidth]{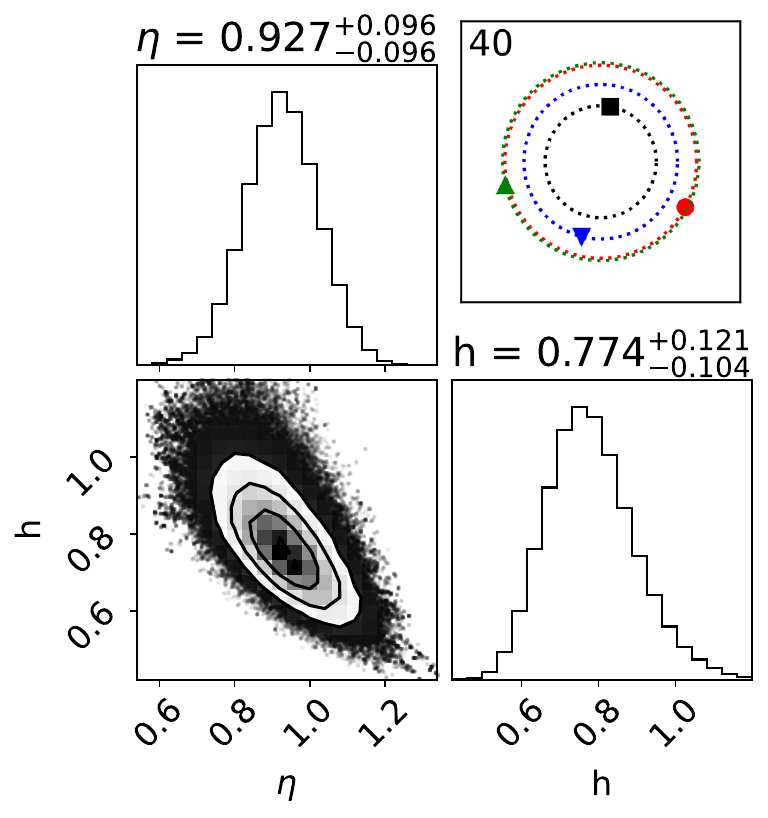}
    \includegraphics[width=0.35\textwidth]{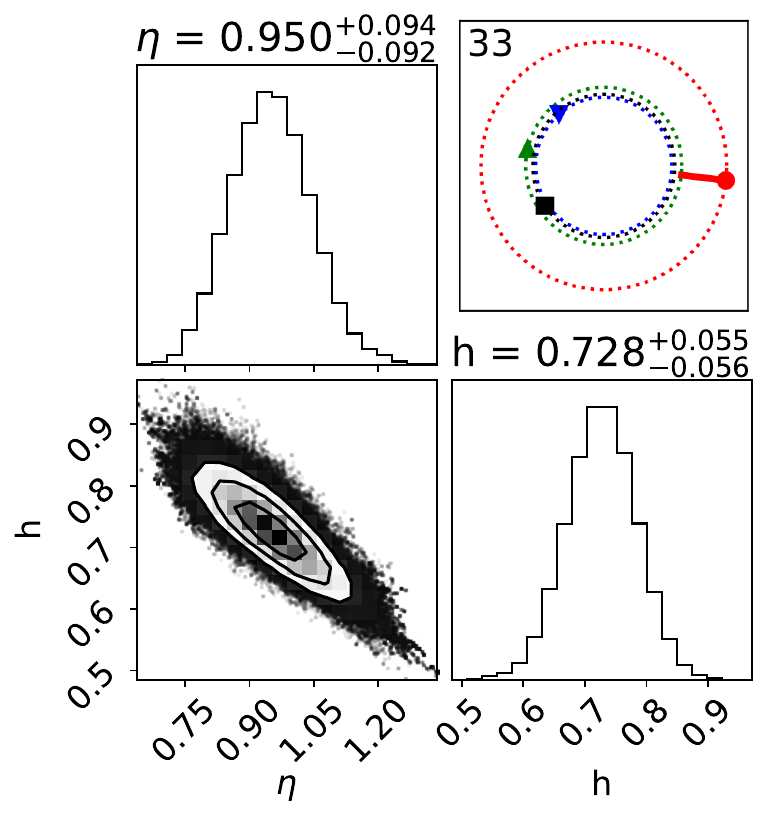}
    \caption{The $(\eta, h)$ MCMC results for EPL+XS Mock 40 \emph{(top)} and Mock 33 \emph{(bottom)}, which have the smallest and largest values of $\Delta r_{1,2}$ in the sample, respectively. These two configurations demonstrate the contrast in constraints based on spacing of the first (red circle) and second (green triangle) arriving images. The annulus length $\Delta r_{1,2}$ is indicated with the red solid line.}
    \label{configs}
\end{figure}

Results for the full sample of 100 mock EPL+XS quads are shown in Figure \ref{hfit-annulus}, and MCMC results for the EPL+XS systems with the smallest (Mock 40) and largest (Mock 33) values of $\Delta r_{1,2}$ are shown in Figure \ref{configs}. The trends are similar to the CPL+XS case. For both mass cases, fold and short-axis cusp configurations tend to be less biased than crosses and long-axis cusps. However, the EPL+XS biases can be even larger and there is more scatter due to the increased phase volume. The key conceptual result is that the shape of the posterior distribution leads to biases in $\eta$ and $h$ for all of the types of systems we have considered, and the biases are larger when the image annulus is narrower.

\subsection{Combining Probability Distributions for the Hubble Constant}

\begin{figure}
    \centering
    \includegraphics[width=0.35\textwidth]{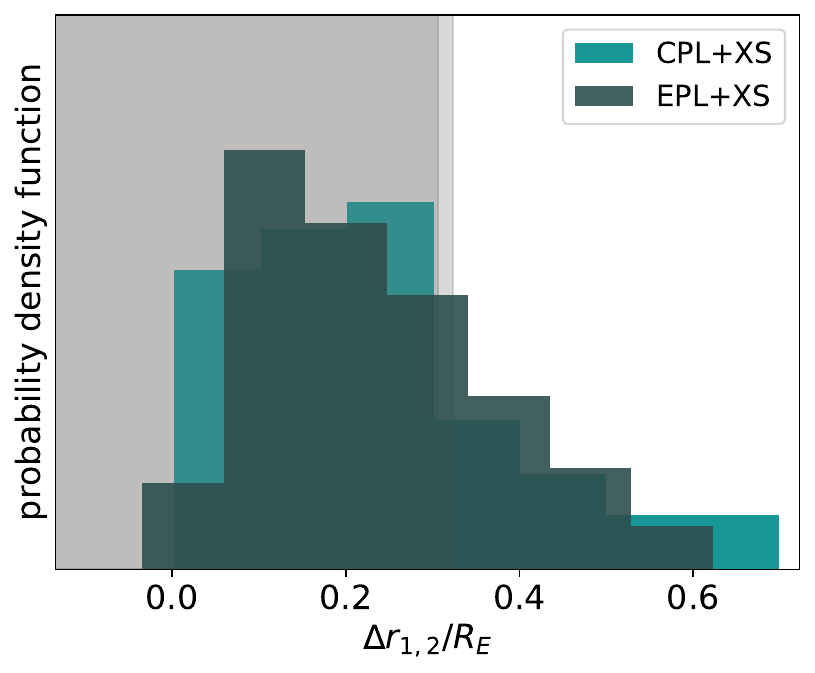}
    \caption{
    Histograms show the distribution of $\Delta r_{1,2} / R_E$ (image annulus widths normalized by the Einstein radius) for the CPL+XS mock lenses {(teal)} and the EPL+XS mock lenses {(dark teal)}. The shaded grey region shows the cutoff at $\Delta r_{1,2}/R_E = 0.307$ for CPL+XS and 0.323 for EPL+XS.
    }
    \label{annulus}
\end{figure}

We now consider joint constraints on $h$ from each ensemble of mock lenses. Given the relation between $h$ bias and annulus length in Figure \ref{hfit-annulus}, we remove more biased systems by imposing a cut based on the annulus length. Such a cut can be replicated in real lens samples since the annulus length is directly observed. Since the bias decreases as annulus length increases, we restrict attention to 25 systems with the highest annulus length. Figure \ref{annulus} shows the cuts for the CPL+XS and EPL+XS systems compared to the annulus length distribution, where 25 of 100 systems remain for each mass case. Although this selection excludes a majority of the original sample, the number of systems is comparable to what has been used in time delay studies to date. For LSST studies, taking 25\% of the full sample will still yield a large ensemble, and will likely be necessary in order to collect all of the ancillary data needed to enable detailed modelling.

\begin{figure}
    \centering
    \includegraphics[width=0.37\textwidth]{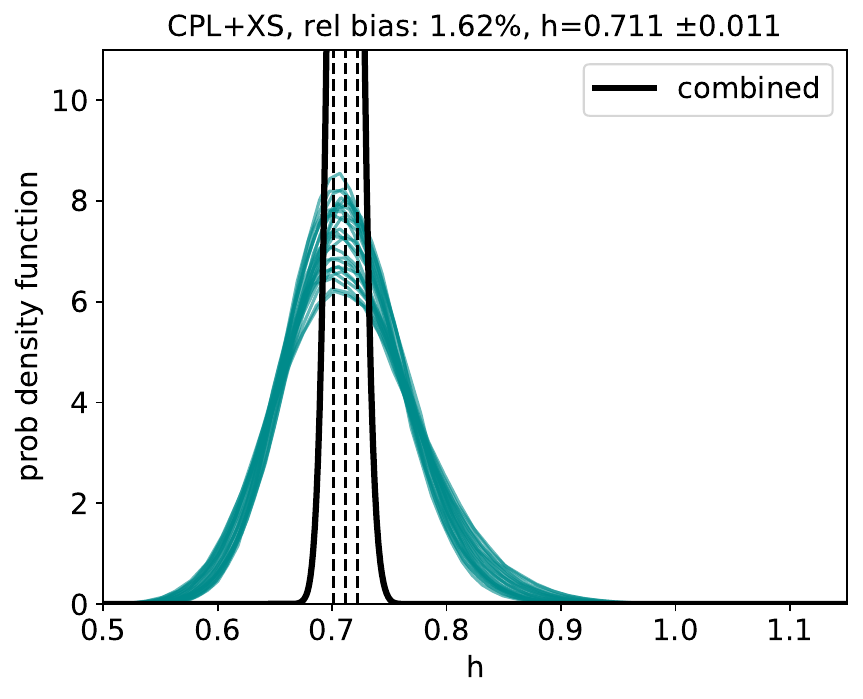}
    \includegraphics[width=0.37\textwidth]{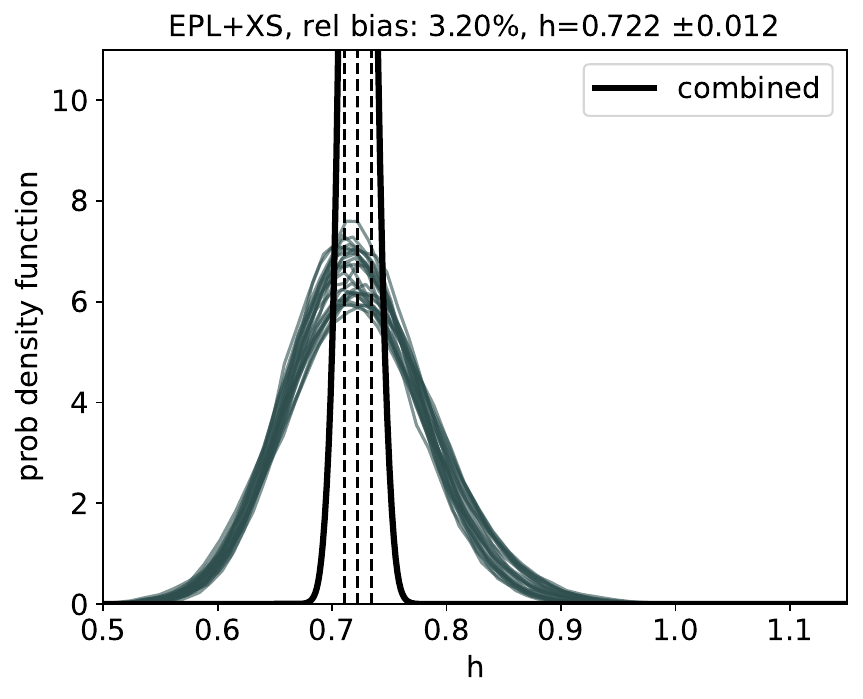}
    \caption{
    After cutting the data by image annulus, we show the probability densities for $h$ given the information from each mock lens system (colored curves) and the overall probability density for $h$ using all the systems plotted in each panel (black curves). \emph{(top)} The CPL+XS lenses exhibit a relative bias of only $1.62\%$, while \emph{(bottom)} the EPL+XS lenses exhibit more relative bias of $3.20\%$. }
    \label{kde-combine}
\end{figure}

For each mock lens, we convert the samples from MCMC into a probability density function using Gaussian kernel density estimation. We treat the mock lenses as statistically independent from each other, so we can combine results by multiplying the probability densities. The overall bias is calculated using the median of the combined distribution, and errors are reported as the 68\% confidence interval. Using all systems for the most likely $h$ results in $h=0.713 \pm 0.007$ (1.86\% bias) for CPL+XS and $h=0.730 \pm 0.008$ (4.22\% bias) for EPL+XS. Figure \ref{kde-combine} shows how the annulus length cut can reduce the bias to $h=0.711 \pm 0.011$ (1.62\% bias) for CPL+XS and $h=0.722 \pm 0.012$ (3.20\% bias) for EPL+XS.

\subsection{Adding Noise}
\label{noise}
\begin{figure}
    \centering
    \includegraphics[width=0.4\textwidth]{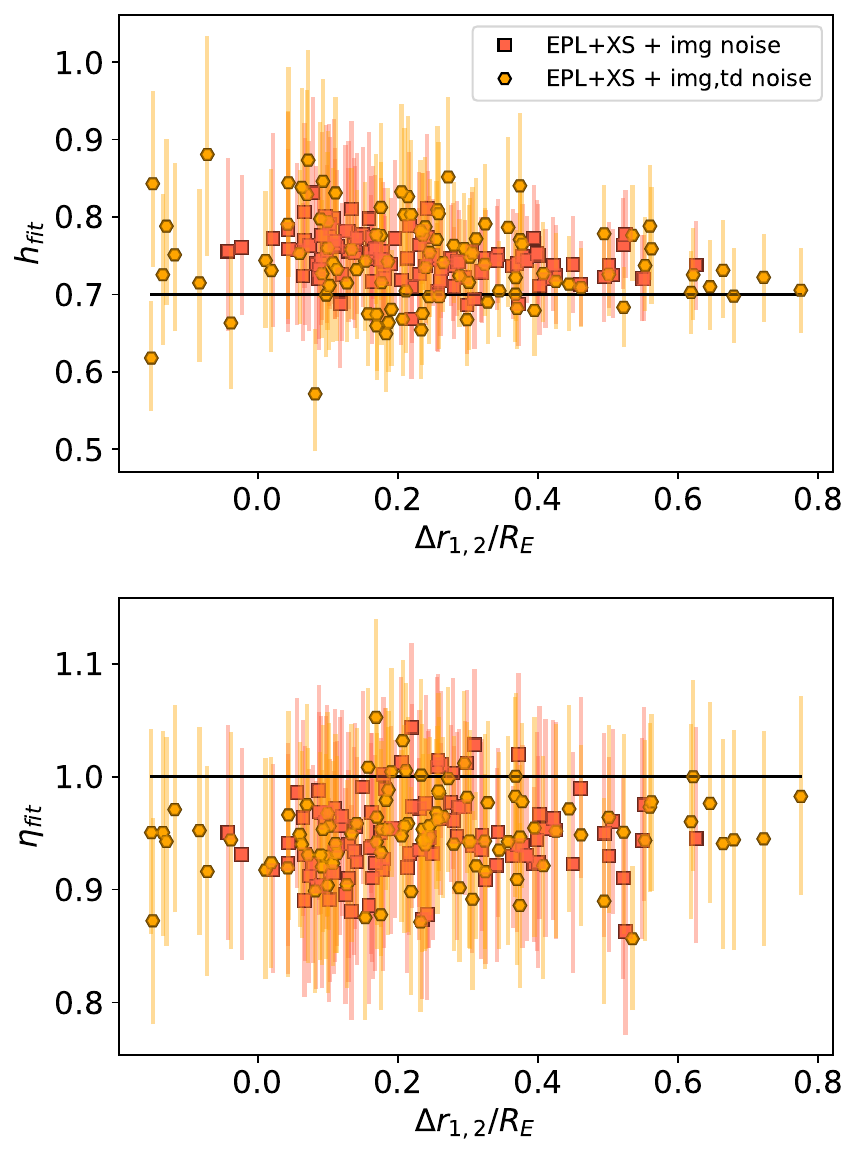}
    \caption{Similar to Figure \ref{hfit-annulus}, but now including noise first in the image positions alone (red), and then image positions and time delays (orange). Some annulus lengths shift due to the image position noise and new time delay order in some cases.}
    \label{eplxsnoise}
\end{figure}

\begin{figure}
    \centering
    \includegraphics[width=0.37\textwidth]{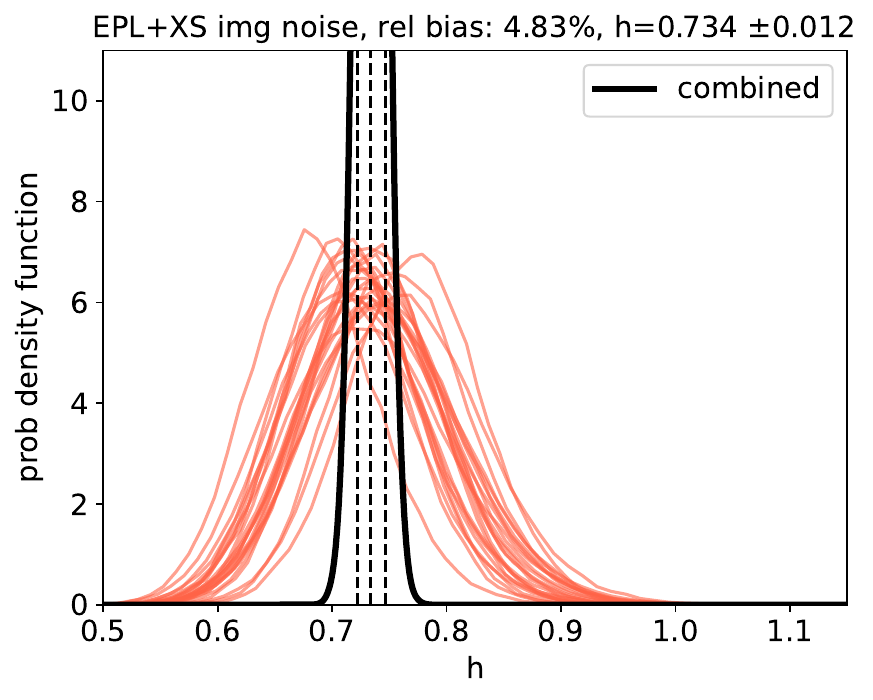}
    \includegraphics[width=0.37\textwidth]{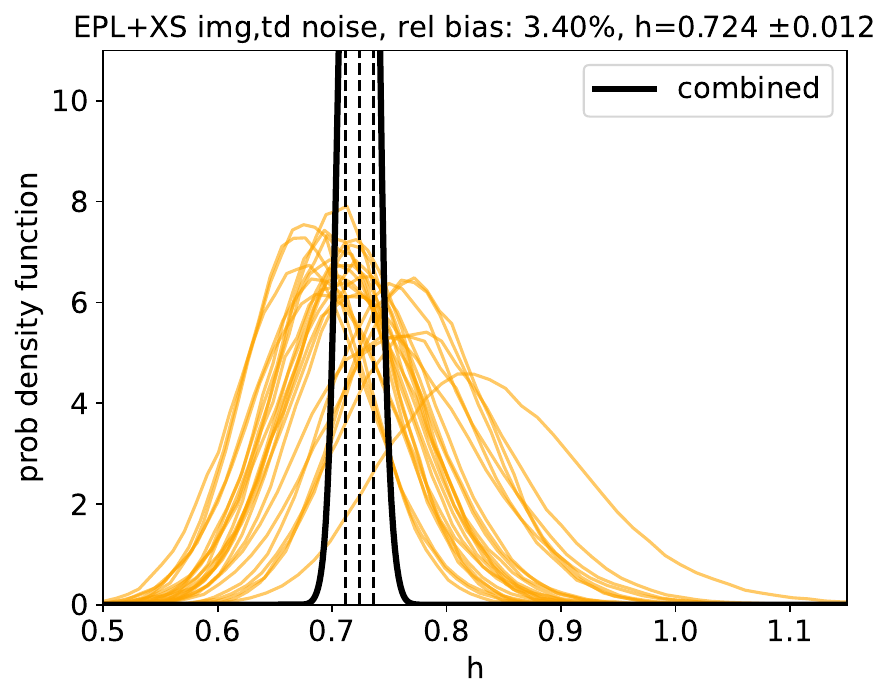}
    \caption{The combined $h$ probability curve for the EPL+XS sample with image position noise \emph{(top)} and image position and time delay noise \emph{(bottom)}. Each case uses systems with $\Delta r_{1,2}$ above $0.320$ arcsec and $0.357$ arcsec, respectively. Although the percent bias is higher with noise, these results are within 1$\sigma$ of the result in Figure \ref{kde-combine}.}
    \label{noisebias}
\end{figure}

We consider what happens if noise is introduced to the image positions, and then to the image positions and time delays. The original EPL+XS sample gets noise randomly drawn from Gaussians with standard deviations of $\sigma_{pos}=0.003$ arcsec and $\sigma_{t} = 2.0$ d. Perturbing by Gaussian noise not only increases the uncertainties, but can also alter the time delay order between images. Figure \ref{eplxsnoise} shows the medians and uncertainties for $h$ and $\eta$ with the image position noise (red), then the image position and time delay noise (orange). The noise introduces scatter that somewhat obscures the trend with image annulus, but the biases in $\eta$ and $h$ are still apparent. If all 100 EPL+XS systems are combined, we find $h=0.735 \pm 0.008$ (4.95\%) for image position noise and $h=0.727\pm 0.008$ (3.88\%) for both image position and time delay noise. Figure \ref{noisebias} shows that selecting systems with the 25 highest $\Delta r_{1,2}$ values results in $h=0.734\pm0.012$ $(4.83\%)$ for image position noise and $h=0.724\pm 0.012$ $(3.40\%)$ for both image position and time delay noise. For comparison, if we select \emph{random} subsamples of 25 systems, the percent bias ranges from 3-7\% for image position noise and 2-9\% for both image position and time delay noise. We conclude that selecting by annulus length still tends to reduce the bias in $h$ and $\eta$ even though measurement noise tends to wash out the trend with image annulus to some degree.

\subsection{Comparing Different Time Delay Errors}

\begin{figure}
    \centering
    \includegraphics[width=0.37\textwidth]{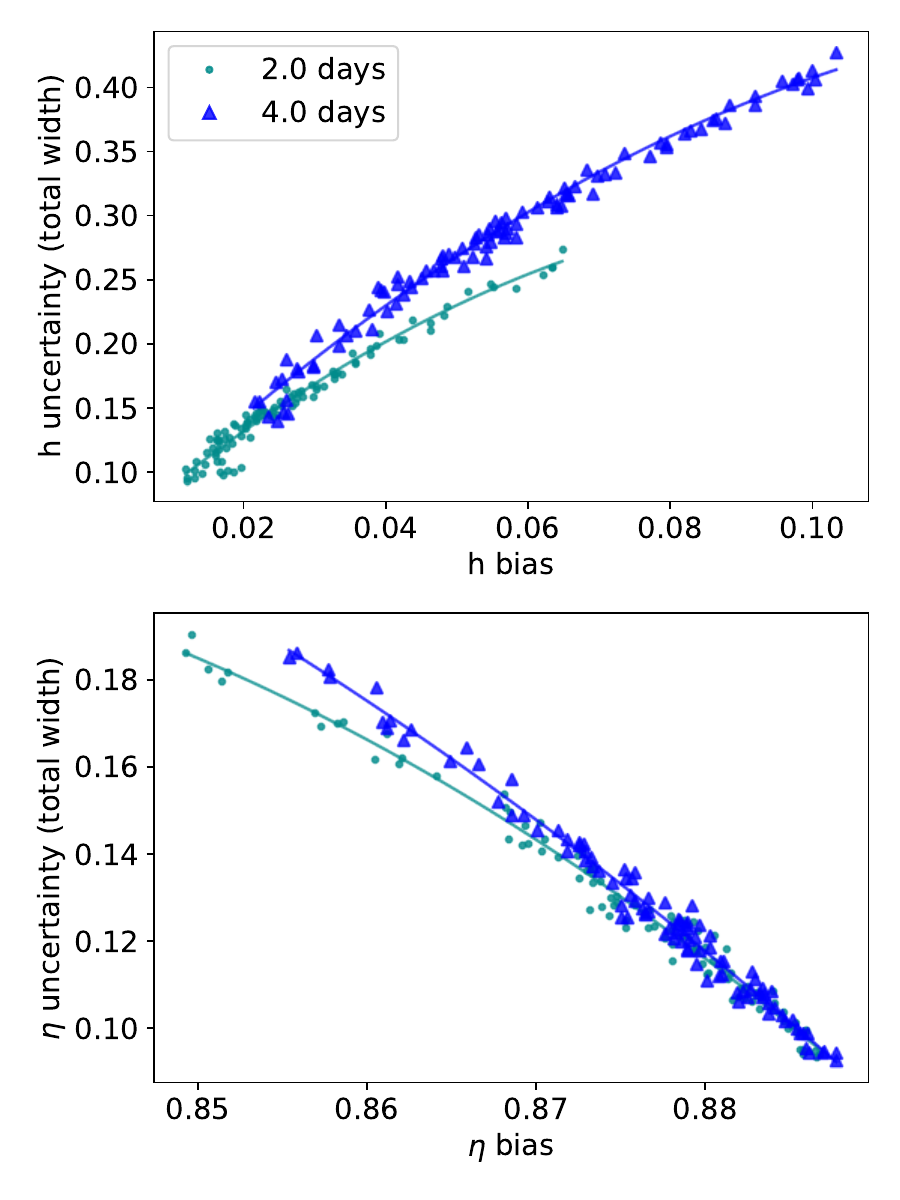}
    \caption{
    Bias and uncertainty for $h$ \emph{(top)} and $\eta$ \emph{(bottom)}, using CPL+XS models. Teal points show our fiducial results with time delay uncertainties of 2 days, while blue points show results with larger time delay uncertainties of 4 days. The curves show quadratic fits to the points (mainly to guide the eye).
    }
    \label{bias-uncertainty}
\end{figure}

From Figure \ref{hfit-annulus}, it is apparent that models with more bias in $h$ also have larger uncertainties; Figure \ref{bias-uncertainty} illustrates that relation for the CPL+XS models. Like the bias itself, the connection between bias and uncertainty arises from the `flared' shape of the posterior distribution for $\eta$ and $h$. In our final test, we double the time delay uncertainties to 4 days (using the same source positions and image configurations as before). Figure \ref{bias-uncertainty} shows that having larger time delay uncertainties increases the bias and uncertainty for $h$ but has little effect on $\eta$. Conceptually, the power law slope $\eta$ is mainly determined by the image positions, while $h$ is sensitive to the time delays.

\section{Conclusions}
\label{conclusions}

Many studies of $H_0$ bias with lensing focus on how discrepancies between the fitted and intrinsic lens mass distributions lead to bias. We have found that even while using the \emph{same} model to generate and fit lens galaxies, a statistical bias can occur at the level of 3-5\%, depending on whether the elliptical lenses have noise or not. The bias is a consequence of the shape of the posterior distribution. The radial profile degeneracy creates a covariance between $h$ and the power law slope $\eta$, and the posterior broadens as $h$ increases and $\eta$ decreases, so there is more phase space at higher $h$ and lower $\eta$. We summarize our key conclusions as follows:
\begin{itemize}

    \item This statistical bias occurs even in a simple mass distribution like the CPL without shear (Fig.~\ref{doub_circ1&2}). When the image annulus is narrow, the images do not provide enough information to constrain the radial mass profile, and the radial profile degeneracy means that bias and uncertainty in $\eta$ translate into bias and uncertainty in $h$.

    \item For the CPL+XS case, the fitted $h$ and $\eta$ values from the MCMC fit follow a trend with the relative time delay or annulus length between the first and second images (Fig.~\ref{hfit-annulus}). For all lens systems, $h$ is consistently overestimated, while $\eta$ is underestimated. The amount of bias is smaller with greater annulus length.

    \item The phase space volume expands with more complicated  mass models, and there is more scatter in the fitted values for the EPL+XS case (Fig.~\ref{hfit-annulus}). The max bias is worse for the EPL+XS case (Fig.~\ref{mock_ellip_maxbias}) as compared to the CPL+XS case (Fig.~\ref{CPLXS-corner}).
    
    \item The trend with annulus length suggests a way to mitigate the bias. If we select systems with annulus lengths $\Delta r_{1,2} \gtrsim 0.3$ arcsec, we find an overall $h$ bias of $1.62\%$ for the CPL+XS mock lenses, and $3.20\%$ and for the EPL+XS case (Fig.~\ref{kde-combine}). The annulus length selection reduces, but does not eliminate, the bias.

    \item Measurement noise introduces scatter that somewhat obscures the relation between image annulus and bias, but we still find that selecting systems with large annuli tends to reduce the bias (Fig.~\ref{noisebias}).
    
    \item The bias and uncertainty are correlated with each other. Although higher $h$ bias is associated with higher $\eta$ bias, the bias and uncertainty for $h$ are more sensitive to time delay errors than for $\eta$ (Fig.~\ref{bias-uncertainty}). 

\end{itemize}

It is not clear how much this statistical bias might affect current lensing measurements of $H_0$. The true mass distribution of lens galaxies are likely to be more complicated than standard lens models, so there may be a mismatch between true and fitted mass distributions that could lead to a systematic bias, different from what we have examined here. Some systems may only have image position and time delay constraints like in our study \citep{2023ApJ...948..115P}, while other systems also have arcs and stellar kinematics \citep{2019MNRAS.490..613S, 2023A&A...673A...9S}. Such information can help constrain the radial mass profile in ways that we have not yet explored. Follow-up work should seek to disentangle the effects of the statistical bias we have examined here from systematic biases associated with model mismatch. Follow-up work could also incorporate more observational effects, or different values of ellipticity, shear, and source positions. Selection effects (like the sample being restricted to quads) could contribute to this bias. This would be corrected for in the likelihood function, and would require additional computational effort for efficient MCMC sampling.

In the era of the Hubble tension and large astronomical surveys, lens samples are growing and follow-up observations are often necessary in order to constrain lens models and $H_0$ with the level of precision necessary to help resolve the Hubble tension. Our results suggest that follow-up observations should prioritize lenses in which the images span wide annuli in order to limit the uncertainties and hence the statistical bias in $\eta$ and $h$.

\section*{Acknowledgements}
We thank Simon Birrer, Anowar Shajib, and Alessandro Sonnenfeld for their helpful suggestions. We also thank the Rutgers lensing group: Lana Eid, Somayeh Khakpash, and Aniruddha Madhava for helpful discussions. This work uses the code \verb|pygravlens|, \verb|lensmodel| \citep{lensmodel}, Python 3 \citep{Python}, \verb|emcee| \citep{emcee}, \verb|numpy| \citep{numpy}, \verb|scipy| \citep{2020SciPy-NMeth}, and \verb|pandas| \citep{2022zndo...6408044R}.

\section*{Data Availability}
The data underlying this article are available in GitHub at \url{https://github.com/dilysruan/paper-ruan-keeton-2023}.

\bibliographystyle{mnras}
\bibliography{bibby_mnras}



\appendix

\section{Modelling choices}
\label{app-choices}

In addition to the mass model, there are other choices that must be made in the modelling process. This work focused on bias with respect to image configurations, and we expect the results to be qualitatively similar if this analysis was performed using a different modelling choice. Still, it is worth quantifying how much these choices affect the results. For illustration, we perform these tests using two mock lenses that exhibit low and high bias with EPL+XS models, namely Mock 4 (a fold configuration) and Mock 51 (a cross configuration). We find that discrepancies between modelling choices is more significant for Mock 51 than for Mock 4. The `control' mode used for the CPL+XS and EPL+XS  in this paper is: marginalized source position, no flux constraints, and the Einstein radius parameterized through $b$.

\begin{figure}
    \centering
    \includegraphics[width=0.47
    \textwidth]{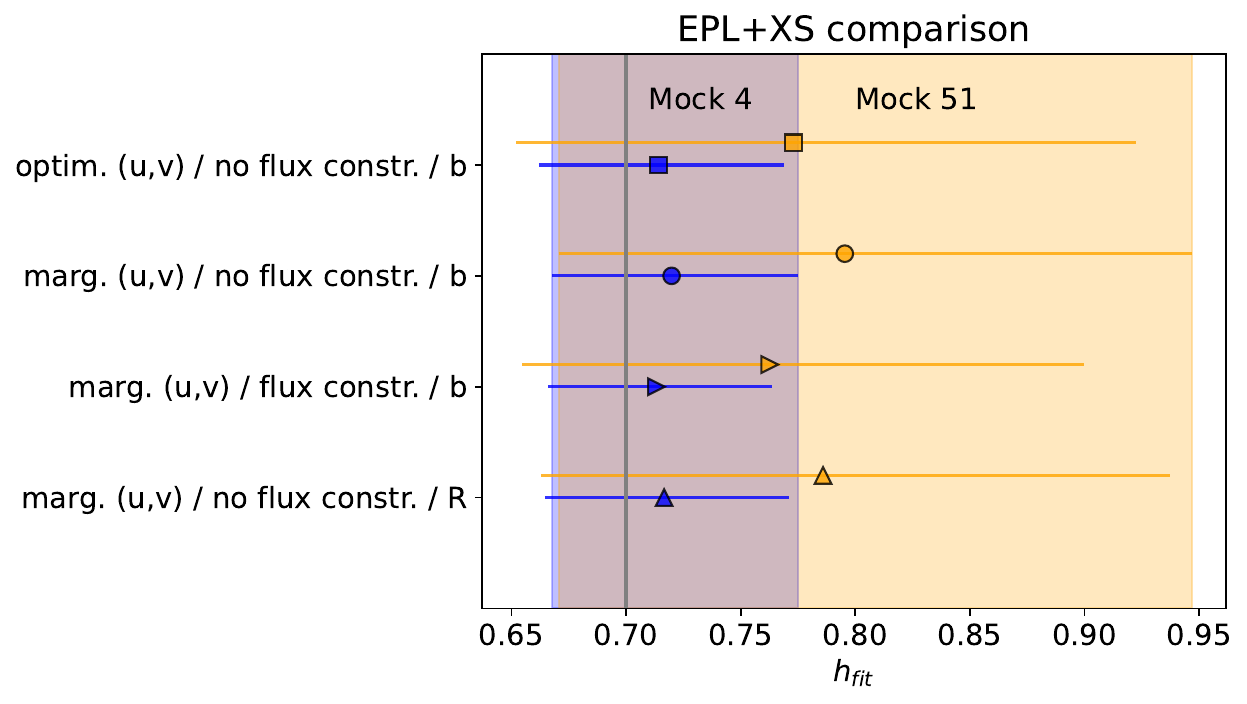}
    \caption{
    Comparison of the median (markers) and 68\% confidence intervals (error bars) for different modelling choices with EPL+XS Mock 4 (blue) and Mock 51 (orange).
    We consider whether the source coordinates are optimized or fully marginalized (rows 1--2), whether to include flux constraints and optimize the source flux (row 3), and whether to normalize using the `R mode' (row 4). The shaded region shows the `control' mode (row 2) constraints.
    The vertical grey line shows the true input $h=0.7$.
    }
    \label{choices}
\end{figure}

As described in Section \ref{algorithm}, our code marginalizes the source coordinates $\boldsymbol{\beta} = (u,v)$ by default. It is computationally simpler to optimize those parameters, so that approach has been adopted in the past (e.g., in \verb|lensmodel|; \citealt{lensmodel}). The comparison between source optimization and marginalization for each mock lens is shown in Figure \ref{choices}, where no flux constraints are used for either case. The $h$ bias is slightly smaller for both mock lenses in the optimization mode. All of the remaining tests use the marginalization mode. 

Image flux ratios provide additional constraints on lens models. We add four more constraint terms (for a quad lens) and introduce one more free parameter, namely the source flux, which we optimize. Figure \ref{choices} shows that using these constraints reduces $h$ bias and uncertainty for both mock lenses. We omit flux constraints in most of our analysis of quad lenses for simplicity, and since flux constraints are susceptible to microlensing.

As shown in Equations (\ref{EPL})--(\ref{bmode}), the lens model can be parameterized using either the Einstein radius $R_E$ (which we call `R mode') or the simple multiplicative factor $b$ (`b mode'). The two parameters are equivalent for a power law with $\eta = 1$, but for $\eta \ne 1$ they slightly differ and the statistical sampling of the parameter space might also differ. As shown in Figure \ref{choices}, the `R mode' results are slightly more accurate than the control.

\begin{figure}
\includegraphics[width=0.37\textwidth]{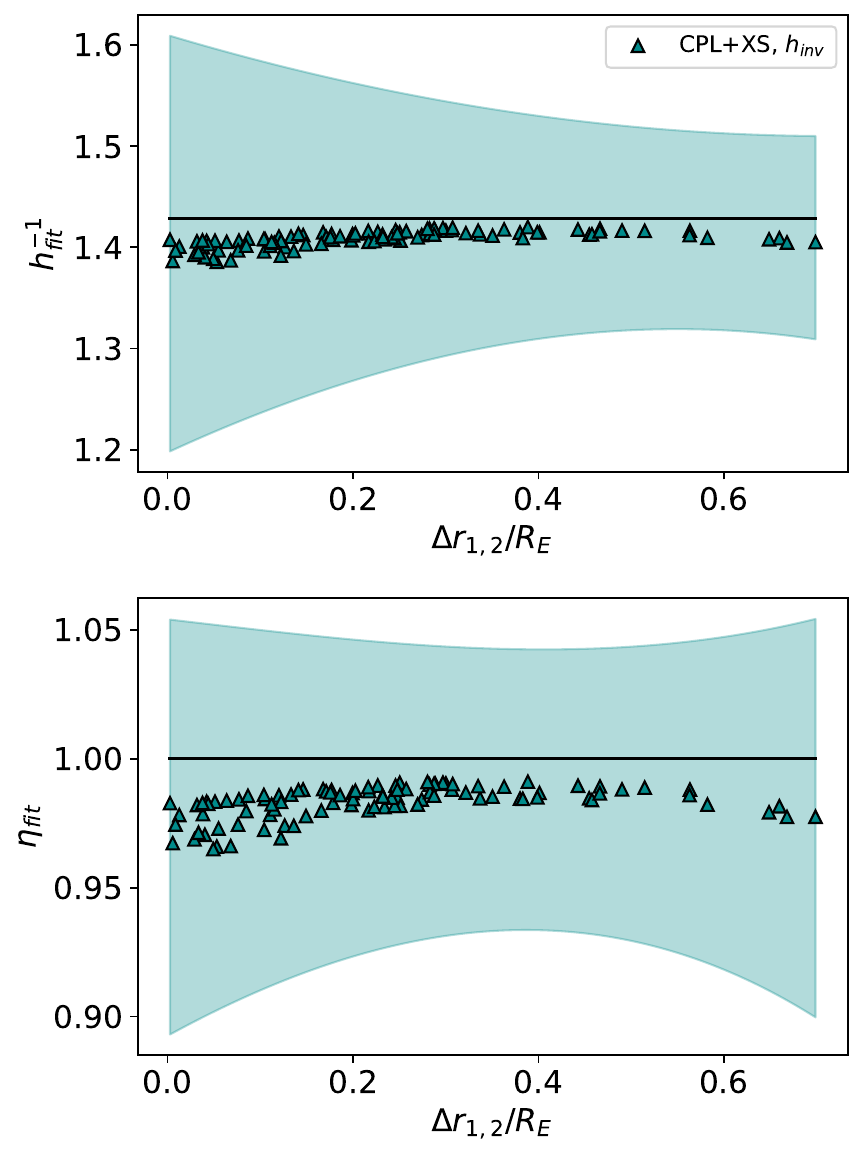}
    \caption{Similar to Figure \ref{hfit-annulus}, the bias as a function of normalized annulus length, but with $h^{-1}$ as the free parameter.
    }
    \label{hinv}
\end{figure}

\begin{figure}
    \centering
    \includegraphics[width=0.37\textwidth]{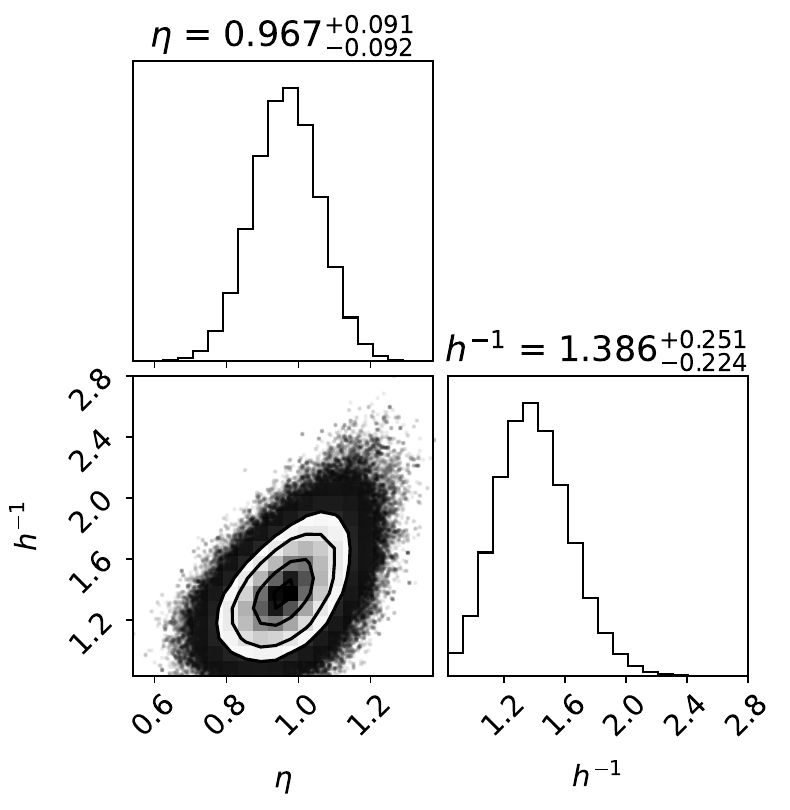}
    \caption{MCMC results for CPL+XS Mock 35, fitting for $h^{-1}$.}
    \label{hinv_4}
\end{figure}

Instead of directly fitting for $h$, its inverse may be used if one fits for the time delay distance, where $D_{\Delta t} \propto h^{-1}$. We analyze how bias manifests with a $h^{-1}$ parameterization for CPL+XS lenses. We saw overestimated $h$ in the main analysis, and we see a bias for underestimated $h^{-1}$. Figure \ref{hinv} shows how both $h^{-1}$ and the mass power law slope are biased low. Figure \ref{hinv_4} shows the phase space contours for CPL+XS Mock 35, in contrast to fitting for $h$ in Figure \ref{CPLXS-corner}.

\section{Corner plots}
\label{app-corner}

\begin{figure*}
    \centering
    \includegraphics[width=0.9\textwidth]{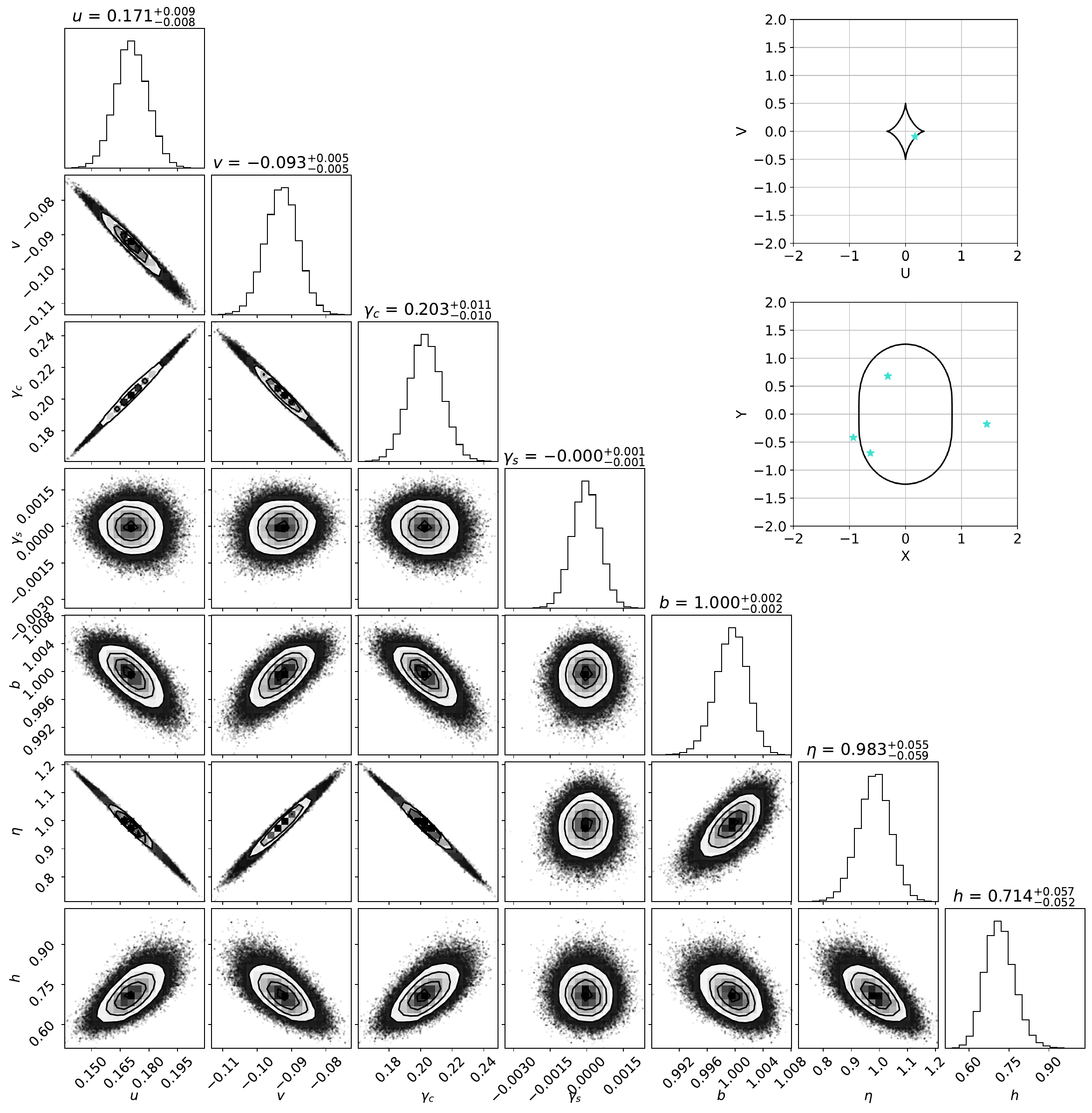}
    \caption{
    MCMC results for CPL+XS Mock 77. Here $u$ and $v$ represent the source position coordinates, $b$ is the Einstein radius parameter, and $\gamma_c$ and $\gamma_s$ are the quasi-Cartesian components of external shear. The posterior distribution is fairly compact and symmetric. There are notable covariances between certain parameters. The upper-right inlay shows the caustics and source position in the source plane (\emph{top}) as well as the critical curves and image positions in the image plane (\emph{bottom}) for this system. Each plane is on an arcsec$^2$ grid.
    }
    \label{mock77-corner}
\end{figure*}

\begin{figure*}
    \centering
    \includegraphics[width=0.9\textwidth]{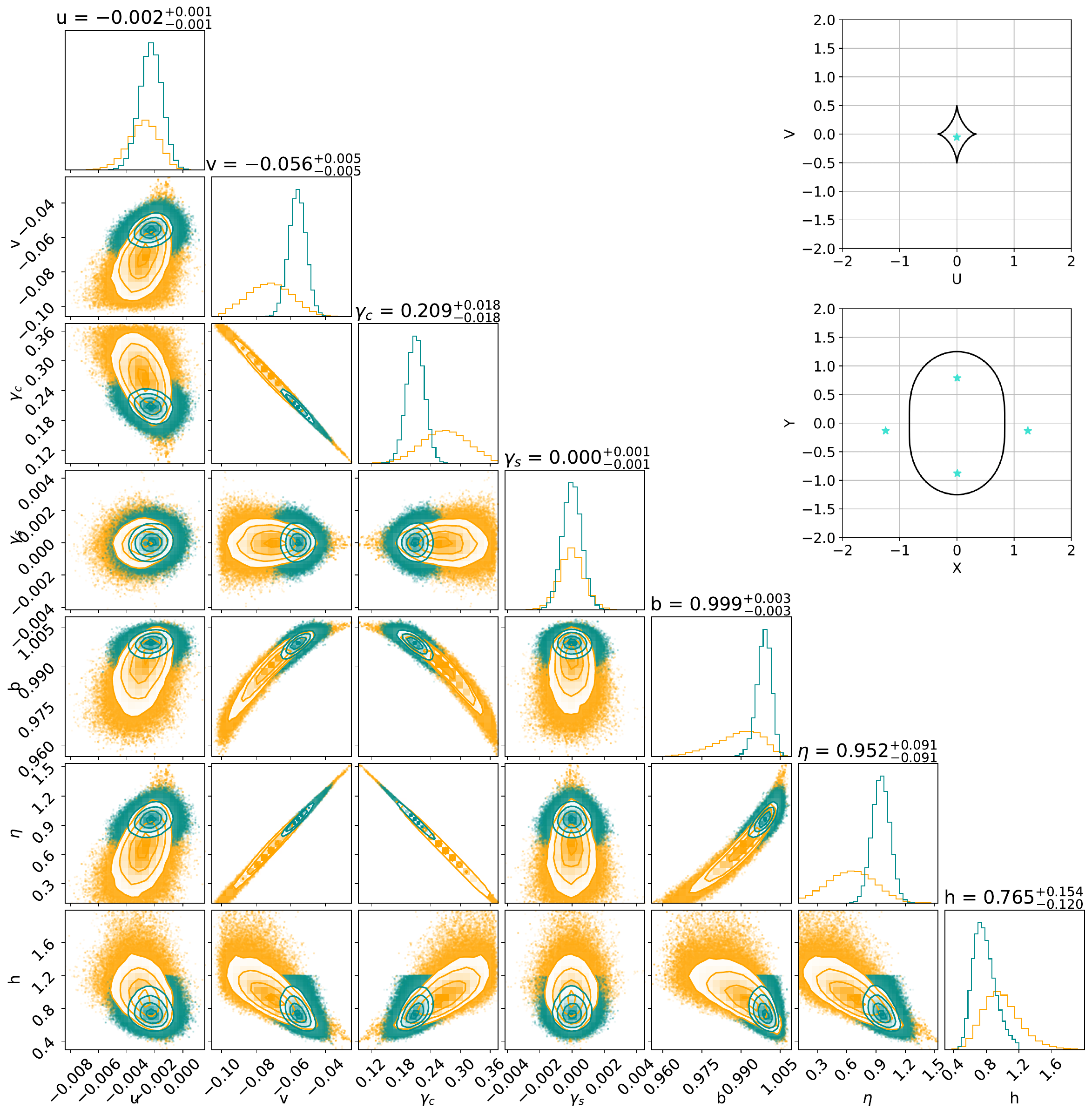}
    \caption{
    MCMC results for CPL+XS Mock 35, an outlier in the sample. Using only image positions and time delays as constraints (orange), the posterior distribution extends slightly beyond our priors for some parameters, resulting in $\eta = 0.648^{+0.242}_{-0.248}$ and $h = 0.972^{+0.248}_{-0.204}$. This symmetric cross configuration is better constrained (teal and column titles) if we apply constraints on $\eta$ through the term $\chi_{PL}^2$ in Equation (\ref{chi_PL}).
    Note: we imposed bounds $\eta \in (0.1,1.9)$ for the orange case and $\eta \in (0.5,1.5)$ and $h\in (0,1.2)$ for the teal case.
    }
    \label{CPLXS-corner}
\end{figure*}

\begin{figure*}
    \centering
    \includegraphics[width=0.99\textwidth]{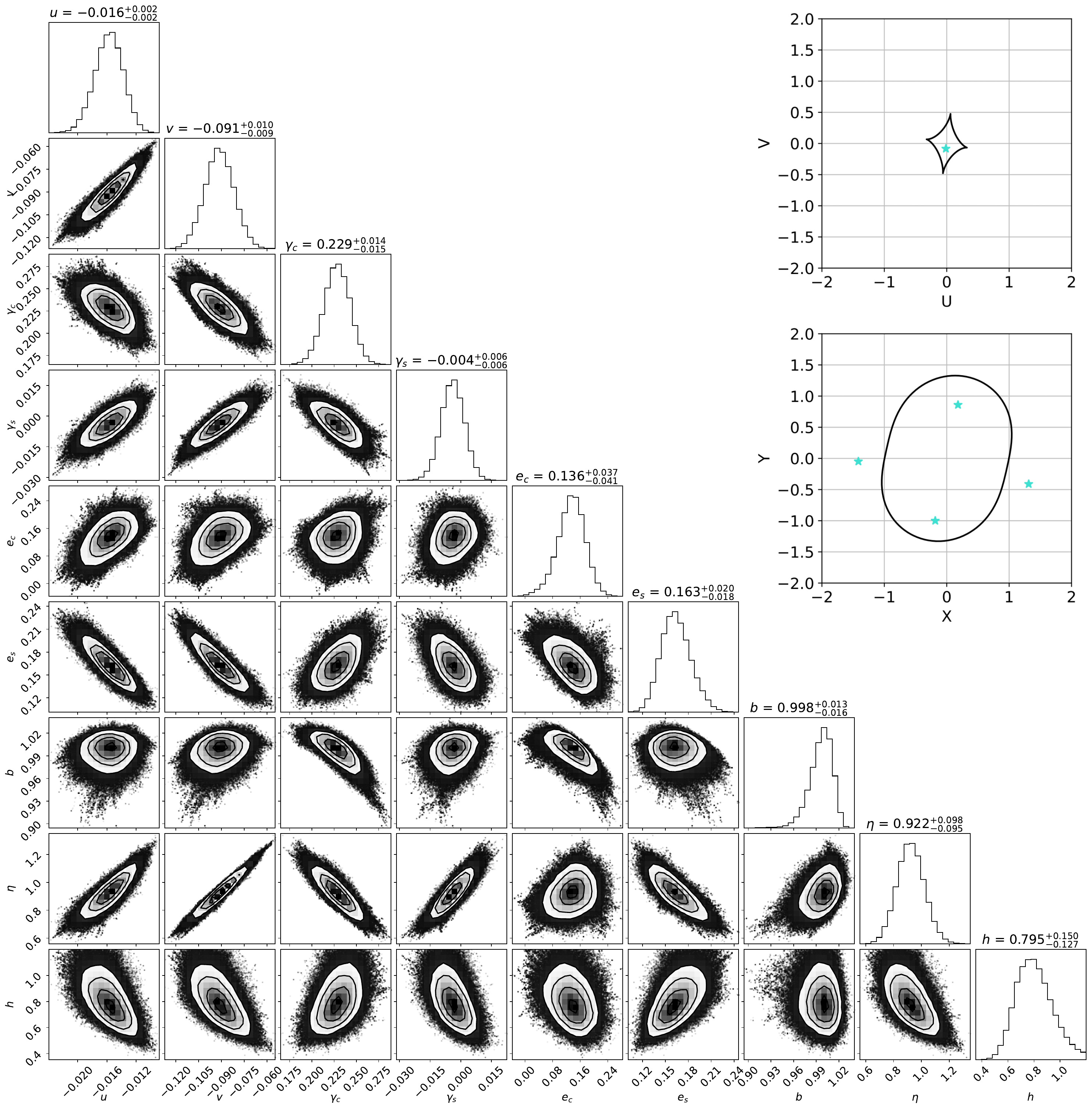}
    \caption{
    MCMC results for EPL+XS Mock 51. In addition to the fitted parameters in Figure \ref{mock77-corner}, this model includes the quasi-Cartesian ellipticity coordinates $e_c$ and $e_s$, which makes the phase space contours more complicated than the CPL+XS systems.
    }
    \label{mock_ellip_maxbias}
\end{figure*}

The full corner plots are shown for a few lenses in Figure \ref{mock77-corner} (CPL+XS Mock 77), Figure \ref{CPLXS-corner} (CPL+XS Mock 35 -- with and without the $\eta$ prior), and Figure \ref{mock_ellip_maxbias} (EPL+XS Mock 51). 

\bsp	
\label{lastpage}
\end{document}